\documentclass[ aps,%
floatfix,%
final,%
notitlepage,%
oneside,%
onecolumn,%
nobibnotes,%
nofootinbib,%
superscriptaddress,%
showpacs,%
]%
{revtex4}

\usepackage{epsfig}
\usepackage{amsfonts}
\usepackage{amsmath}
\usepackage{epsfig}
\usepackage{graphics}
\usepackage{axodraw}

\newcommand{\beq}{\begin{eqnarray}}\newcommand{\eeq}{\end{eqnarray}}
\newcommand{\beqa}{\begin{eqnarray*}}\newcommand{\eeqa}{\end{eqnarray*}}

\begin{document}

\title{Decay $\eta_b \to J/\psi J/\psi$ in light cone formalism}
\author{V.V. Braguta}
\email{braguta@mail.ru}
\affiliation{Institute for High Energy Physics, Protvino, Russia}
\author{Kartvelishvili, V.}\email{V.Kartvelishvili@lancaster.ac.uk}
\affiliation{Lancaster University, Lancaster, UK}

\begin{abstract}
The decays of pseudoscalar bottomonium $\eta_b $ into a pair of vector 
charmonia, $J/\psi J/\psi, J/\psi \psi', \psi' \psi'$ are considered in the 
light cone formalism. 
Relativistic and leading logarithmic radiative corrections to the amplitudes 
of these processes are resummed. It is shown that the small value for the 
branching ratio of the decay $\eta_b \to J/\psi J/\psi$ obtained within the leading 
order nonrelativistic QCD is a consequence of a fine-tuning between 
certain parameters, which is broken when relativistic and leading logarithmic 
radiative corrections are taken into account. As a result, the branching ratio 
obtained in this paper is enhanced by an order of magnitude.
\end{abstract}


\maketitle

\newcommand{\ins}[1]{\underline{#1}}
\newcommand{\subs}[2]{\underline{#2}}
\vspace*{-1.cm}
\section{Introduction}

Ever since the discovery of the $\Upsilon$ meson, there have been numerous attempts
of observing the lightest pseudoscalar bottomonium state, $\eta_b$. 
However, only recently the first experimental evidence of the existence of 
this meson was found by BaBar collaboration, in the radiative decay 
$\Upsilon(3S) \to \eta_b + \gamma$ \cite{:2008vj}. 
Its mass was found to be $m_{\eta_b}=9388^{+3.1}_{-2.3}(stat)\pm 2.7(syst)$~MeV,
but our knowledge of its other properties remains rather poor. 

In \cite{Braaten:2000cm} it was proposed to look for the $\eta_b$ meson 
in the decay $\eta_b \to J/\psi J/\psi$, but, despite its clean signature, 
this process may be hard to observe due to its extremely small branching ratio:
contrary to other similar processes, such as 
the decays $\chi_b \to J/\psi J/\psi$ \cite{Kartvelishvili:1984en}, 
the rate of the decay $\eta_b \to J/\psi J/\psi$ vanishes 
at the leading order of both relative velocity and $1/M_{\eta_b}$ expansions.
The calculations made within nonrelativisitic QCD (NRQCD) \cite{Bodwin:1994jh}  
yield $Br(\eta_b \to J/\psi J/\psi) \sim 10^{-8}-10^{-7}$ \cite{Jia:2006rx, Gong:2008ue},
however in \cite{Santorelli:2007xg} it was 
shown that the account of final-state interaction effects can enhance 
it up to about $10^{-5}$.

A similar conclusion can be drawn from the comparison of the decays 
$\eta_b \to J/\psi J/\psi, J/\psi \psi', \psi' \psi'$ 
and the processes of double charmonia production at B-factories.
It is now clear that these processes are greatly effected by radiative 
and relativistic corrections 
\cite{Braaten:2002fi, Liu:1, Liu:2, Zhang:2005ch, Gong:2007db, Zhang:2008gp,  Bondar:2004sv, Braguta:2005kr, 
Berezhnoy:2007sp, Ebert:2008kj, He:2007te, Bodwin:2007ga, Braguta:2008tg}. 
With the mass of $\eta_b$ being so close to the energy at which B-factories operate,
it is natural to expect that the same is true for the decays 
$\eta_b \to J/\psi J/\psi, J/\psi \psi', \psi' \psi'$, and hence
the consideration of these processes without accounting for radiative and 
relativistic corrections is unreliable. This was also confirmed by the 
calculation of radiative corrections within NRQCD, performed in \cite{Gong:2008ue}.

In this paper, the processes $\eta_b \to J/\psi J/\psi, J/\psi \psi', \psi' \psi'$ are 
considered within the light cone (LC) formalism \cite{Chernyak:1983ej}. In this approach, 
the amplitudes of these processes are expanded in $(M_{c \bar c}/M_{b \bar b})^2 \sim 0.1$, 
which is sufficiently small for the applicability of the method \cite{Braguta:2009df}. 

In the LC formalism, the amplitude of a process under study is decomposed into 
the perturbative part, dealing with the production of quarks and gluons at small distances,
and the large-distance part describing the hadronization of the partons.  
For hard exclusive processes, the latter can be
parameterized by the process-independent distribution amplitudes (DA), 
which can be considered as hadrons' wave functions at lightlike separations 
between the partons inside the hadron.
It should be noted that DAs contain information about the structure of mesons 
and effectively resum relativistic corrections to the amplitude. Moreover, using 
renormalization group evolution of DAs, one can take into account 
the leading logarithmic radiative corrections to the amplitude. 


This paper is organized as follows. In the next section DAs for charmonium are
defined, and various models for these DAs are discussed. In the third section,
the amplitude of the decay of $\eta_b$ into two vector mesons is derived. 
Finally, in the last section the numerical results and their uncertainties are 
presented and discussed.

\section{Distribution amplitudes for charmonium}

The amplitude of the process $\eta_b \to V_1 V_2$, with $V_{1,2}$ standing for 
either $J/\psi$ or $\psi'$, can be parameterized with a single formfactor $F$:
\beq
M=F e_{\mu \nu \sigma \rho} p_1^{\mu} p_2^{\nu} \epsilon_1^{\sigma} \epsilon_2^{\rho},
\label{amp}
\eeq
where $p_1, p_2$ and $\epsilon_1, \epsilon_2$ are the momenta and polarization vectors
of $V_1$ and $V_2$ respectively.
Hence, the width of the decay $\eta_b \to V_1 V_2$ can be written in the form 
\beq
\Gamma[\eta_b \to V_1 V_2]=|F|^2 \frac {|{\bf p}|^3} {4 \pi},
\label{gamma}
\eeq
where ${\bf p}$ is the 3-momentum of a final meson in
the $\eta_b$ rest frame. If the final mesons are identical, $V_1=V_2$,
the width $\Gamma$ should be divided by $2!$.

In the LC  formalism, 
the amplitude of a hard exclusive process
is expanded in the inverse powers of the hard energy scale $E_h$, which 
for the decay $\eta_b \to V_1 V_2$ can be identified as $M_{\eta_b}$. 
The leading order contribution in this expansion requires the two vector mesons 
to be produced with polarizations $\lambda_1=\lambda_2=0$
\cite{Chernyak:1983ej}, but in this case the aplitude (\ref{amp}) vanishes.
In order to obtain a non-zero result, both
vector mesons need to be transversely polarized,
which in turn means that 
the helicities of the quarks in both mesons must be flipped 
twice, and hence leads to a suppression  factor $\sim 1/(M_{\eta_b})^2$ \cite{Jia:2006rx}.
Therefore, the decay $\eta_b \to V_1 V_2$ is 
a next-to-next-to-leading (NNLO) twist process, and in order 
for the calculations to be consistent
one needs DAs up to twist-4. In general, twist-4 DAs 
should contain terms corresponding to higher Fock states 
in addition to the ``valence'' charm quark-antiquark state, 
but we expect such higher states in charmonium to be suppressed,
and in the following we will neglect their contribution.


The DAs for a vector meson $V$ with momentum $p$ and polarization vector $\epsilon$
can be defined as follows
\cite{Ball:1998sk}:
\beq
\langle V(p, \epsilon) | \bar c(x) \gamma_{\rho} [x,-x] c(-x) |0 \rangle
&=& 
f_V M_V \biggl [ 
\frac {(\epsilon x)} {(px)} p_{\rho} \int_{-1}^{1} d \xi e^{i \xi (px)} \bigl ( 
\varphi_1(\xi, \mu) + \frac {M_V^2 x^2} 4 \varphi_2(\xi, \mu)
\bigr ) \nonumber \\ 
&+&\bigr ( \epsilon_{\rho} -p_{\rho} \frac {(\epsilon x)} {(px)} \bigl ) 
\int_{-1}^1 d \xi e^{i \xi (px)} \varphi_3 (\xi, \mu) \nonumber \\
&-&
 \frac 1 2 x_{\rho} \frac {(\epsilon x)} {(px)^2} M_{V}^2 \int_{-1}^{1} d \xi e^{i \xi (px)} \varphi_4 (\xi, \mu)
\biggr ], \nonumber \\ 
\langle V(p, \epsilon) | \bar c(x) \sigma_{\rho \lambda} [x,-x] c(-x) |0 \rangle
&=& 
f_T(\mu)  \biggl [ 
\bigl ( \epsilon_{\rho} p_{\lambda}- \epsilon_{\lambda} p_{\rho} 
\bigr ) \int_{-1}^{1} d \xi e^{i \xi (px)} \bigl ( 
\chi_1(\xi, \mu)   + \frac {M_V^2 x^2} 4 \chi_2(\xi, \mu)
\bigr )  \nonumber \\   
&+&\bigr ( p_{\rho} x_{\lambda}- p_{\lambda} x_{\rho}   \bigl ) \frac {(\epsilon x)} {(px)^2} 
M_V^2 \int_{-1}^1 d \xi e^{i \xi (px)} \chi_3 (\xi, \mu) 
\label{das}
\\
&+& 
\frac 1 2 \bigl ( \epsilon_{\rho} x_{\lambda}- \epsilon_{\lambda} x_{\rho} 
\bigr ) \frac {M_V^2} {(px)}
\int_{-1}^{1} d \xi e^{i \xi (px)} \chi_4 (\xi, \mu) \biggr ],\nonumber \\ \nonumber
\langle V(p, \epsilon) | \bar c(x) \gamma_{\rho} \gamma_5 [x,-x] c(-x) |0 \rangle
&=& 
f_A(\mu) e_{\rho \lambda \alpha \beta} \epsilon^{\lambda} p^{\alpha} x^{\beta} \int_{-1}^1 d \xi e^{i \xi (px)} \Phi_1(\xi, \mu), \nonumber \\
\langle V(p, \epsilon) | \bar c(x)  [x,-x] c(-x) |0 \rangle
&=& 
-i f_S(\mu) (\epsilon x)  \int_{-1}^1 d \xi e^{i \xi (px)} \Phi_2(\xi, \mu). \nonumber 
\eeq
Here $[x,-x]$ is the gluon string which makes the matrix element gauge invariant, 
$\xi$ is a dimensionless variable describing the relative motion of the
charmed quark and antiquark inside the meson,
$\mu$ is the energy scale at which the DAs are defined,  while 
the constants $f_V$ and $f_T(\mu)$ are defined by
\beq
\langle V(p, \epsilon) | \bar c(0) \gamma_{\mu} c(0) |0 \rangle &=& f_V M_V \epsilon_{\mu}, \nonumber \\ 
\langle V(p, \epsilon) | \bar c(0) \sigma_{\mu \nu} c(0) |0 \rangle &=& f_T(\mu) \bigl ( \epsilon_{\mu} p_{\nu}- \epsilon_{\nu} p_{\mu} 
\bigr ).
\eeq
The constants $f_A(\mu), f_S(\mu)$ can be expressed through $f_V, f_T$ as follows:
\beq
f_A(\mu)&=& \frac 1 2 \biggl ( f_V - f_T(\mu) \frac {2 m_c(\mu)} {M_V}  \biggr )M_V , \nonumber \\
f_S(\mu)&=&  \biggl ( f_T(\mu) - f_V \frac {2 m_c(\mu)} {M_V}  \biggr ) M_V^2,
\label{fafs}
\eeq
where $m_c(\mu)$ is the running mass of the $c$ quark.

Eqs. (\ref{das}) contain 10 independent DAs, but 
only 4 of these are relevant for the calculation of the $\eta_b \to V_1 V_2$ decay rate:
$\varphi_1(\xi), \chi_1(\xi), \Phi_1(\xi)$ and $\Phi_2(\xi)$ (see below). 
For the first two, $\varphi_1(\xi)$ and $\chi_1(\xi)$,  we will use models  
proposed in \cite{Braguta:2006wr, Braguta:2007fh, Braguta:2007tq, {Braguta:2008qe}}. 
In \cite{Braguta:2008tg} it was shown that, if the higher Fock states are ignored, 
the functions $\Phi_1(\xi)$ and $\varphi_3(\xi)$ can be unambiguously determined
from the equations of motion. The same is true for the functions $\Phi_2(\xi)$ and $\chi_3(\xi)$. 

In the remainder of this section, a relation between $\Phi_2(\xi), \chi_3(\xi)$ 
and $\varphi_1(\xi), \chi_1(\xi)$ will be derived. 
The functions $\Phi_2(\xi)$ and $\chi_3(\xi)$ can be expanded into a series of Gegenbauer 
polynomials \cite{Ball:1998sk}:
\beq
\chi_3 (x, \mu) &=& \frac 1 2  \biggl [  1 + \sum_{n=2,4..} c_n(\mu) C_n^{1/2} ( 2 x-1 ) \biggr ], \nonumber \\
\Phi_2 (x, \mu) &=& \frac 3 4 (1-\xi^2) \biggl [  1 + \sum_{n=2,4..} d_n(\mu) C_n^{3/2} ( 2 x-1 ) \biggr ].
\eeq
The coefficients $c_n(\mu)$ and $d_n(\mu)$ are related to the moments
of the functions $\varphi_1(\xi), \chi_1(\xi)$ through the equations 
of motion \cite{Ball:1998sk},
\beq
\frac {n+2} 2\langle \xi^n \rangle_{\chi} &=&  \langle \xi^n \rangle_{T} + \frac {n(n-1)} {2} (1-\delta(\mu)) 
\langle \xi^{n-2} \rangle_{\Phi},  \nonumber \\
 (n+1 ) (1-\delta(\mu)) \langle \xi^n \rangle_{\Phi} &=&  \langle \xi^n \rangle_{\chi} -
\delta(\mu) \langle \xi^{n} \rangle_{L},
\label{eom11}
\eeq
where $\langle \xi^n \rangle_{L,T,\chi, \Phi}$ denote the moments of the DAs 
$\phi_1(\xi), \chi_1(\xi), \chi_3(\xi), \Phi_2(\xi)$ respectively, while 
$\delta(\mu)=2 f_V/f_T(\mu) (m_c(\mu) /M_V)$. By solving eqs. (\ref{eom11})
recursively, one can determine the functions $\Phi_2(\xi)$ and $\chi_3(\xi)$. 
In \cite{Braguta:2006wr} it was shown, that there is a fine-tuning of the 
coefficients of the Gegenbauer expansion at the scale 
$\mu \sim \overline{m}_c\equiv m_c(\mu=m_c)$. Without this fine-tuning 
the DAs of a nonrelativistic system would show an unphysical relativistic tail 
already at the scale $\mu \sim \overline{m}_c$. In order to get rid of this tail 
in the DAs $\Phi_2(\xi)$ and $\chi_3(\xi)$, fine-tuning is required between 
the coefficients $c_n, d_n$ and the parameter $\delta$, which
is related to the wave functions $\phi_1(\xi), \chi_1(\xi)$ \cite{Braguta:2008tg}:  
\beq
\delta( \overline{m}_c) = \frac { \int_{-1}^1 \frac {d \xi} {1-\xi^2 } 
\chi_1 (\xi,  \mu \sim \overline{m}_c )} 
{\int_{-1}^1   \frac {d \xi} {(1-\xi^2)^2 }  
\varphi_1  ( \xi, \mu \sim \overline{m}_c ) }.
\eeq

\section{The amplitude of the process $\eta_b \to V_1 V_2$}

The diagrams that contribute to the amplitude of the process 
under study at the leading order in the $\alpha_s$ expansion are shown in Fig. 1.
\begin{figure}[h.t.b]
\begin{picture}(200,100)(0,0)
\SetScale{0.6}
\SetWidth{1.5}
\ArrowLine(50,100)(100,150) \ArrowLine(100,150)(100,50) \ArrowLine(100,50)(50,100)
\SetWidth{0.3}
\Line(20,98)(50,98) \Line(20,102)(50,102)
\Gluon(100,150)(140,150){4}{4} \Gluon(140,50)(100,50){4}{4}
\ArrowLine(140,150)(190,150) \ArrowLine(190,150)(165,100) \ArrowLine(165,100)(140,50)
\ArrowLine(140,50)(190,50) \ArrowLine(190,50)(165,100) \ArrowLine(165,100)(140,150)
\CCirc(50,100){6}{0}{3} \CCirc(190,150){6}{0}{3} \CCirc(190,50){6}{0}{3}
\Line(190,48)(220,48) \Line(190,52)(220,52) \Line(190,148)(220,148) \Line(190,152)(220,152)
\Vertex(100,150){2.5} \Vertex(100,50){2.5} \Vertex(140,150){2.5} \Vertex(140,50){2.5}
\Text(20,70)[]{$\eta_b$}\Text(128,81)[]{$V_1$}\Text(128,38)[]{$V_2$}
\Text(40,81)[]{$b$}\Text(40,38)[]{${\overline{b}}$}\Text(100,36)[]{$c$}\Text(100,84)[]{$c$}
\end{picture}
\begin{picture}(200,100)(0,0)
\SetScale{0.6}
\SetWidth{1.5}
\ArrowLine(100,50)(100,150)\ArrowLine(100,150)(50,100)\ArrowLine(50,100)(100,50)
\SetWidth{0.3}
\Line(20,98)(50,98)\Line(20,102)(50,102)
\Gluon(100,150)(140,150){4}{4}\Gluon(140,50)(100,50){4}{4}
\ArrowLine(140,150)(190,150)\ArrowLine(190,150)(165,100)\ArrowLine(165,100)(140,50)
\ArrowLine(140,50)(190,50)\ArrowLine(190,50)(165,100)\ArrowLine(165,100)(140,150)
\CCirc(50,100){6}{0}{3}\CCirc(190,150){6}{0}{3}\CCirc(190,50){6}{0}{3}
\Line(190,48)(220,48)\Line(190,52)(220,52)\Line(190,148)(220,148)\Line(190,152)(220,152)
\Vertex(100,150){2.5}\Vertex(100,50){2.5}\Vertex(140,150){2.5}\Vertex(140,50){2.5}
\Text(20,70)[]{$\eta_b$}\Text(128,81)[]{$V_1$}\Text(128,38)[]{$V_2$}
\Text(40,38)[]{$b$}\Text(40,81)[]{${\overline{b}}$}\Text(100,36)[]{$c$}\Text(100,84)[]{$c$}
\end{picture}
\caption{The diagrams contributing to the amplitude of the process $\eta_b \to J/\psi J/\psi$
at the leading order in $\alpha_s$.}
\end{figure}
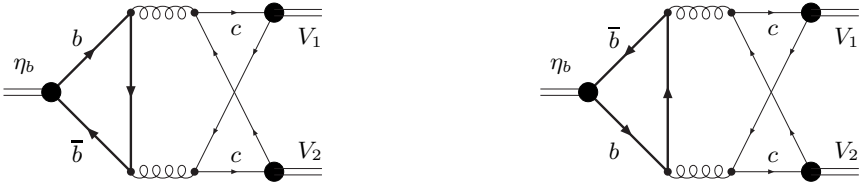
The procedure of calculating the amplitude is described in detail in 
\cite{Chernyak:1983ej}.
This is a lengthy but straightforward exercise, yielding
a result which looks remarkably simple:
\beq
F&=&\int d \xi_1 d \xi_2 H(\xi_1,\xi_2, \mu) \biggl ( 
f_{V1} f_{A2} (\mu) M_{V1} \varphi_1 (\xi_1, \mu) \Phi_1 (\xi_2, \mu) +
f_{V2} f_{A1} (\mu) M_{V2} \varphi_1 (\xi_2, \mu) \Phi_1 (\xi_1, \mu) \nonumber \\
&+&f_{S1}(\mu) f_{T2}(\mu) \chi_1 (\xi_2, \mu) \Phi_2 (\xi_1, \mu) +
f_{S2}(\mu) f_{T1}(\mu) \chi_1 (\xi_1, \mu) \Phi_2 (\xi_2, \mu) \biggr ).
\label{formfactor}
\eeq
Here the function $H(\xi_1, \xi_2, \mu) $ represents the hard part of the amplitude, 
\beq
H(\xi_1, \xi_2, \mu)=\frac {1024 \pi^2 \alpha_s^2(\mu)} {27} f_{\eta_b} \frac 1 {M_{\eta_b}^6}
\frac {1} {(1-\xi_1^2) (1-\xi_2^2) (1+\xi_1 \xi_2) },
\eeq
with the decay constant $f_{\eta_b}$ defined by
\beq
\langle 0 | \bar b(0) \gamma_{\rho} \gamma_5 b(0) |\eta_b(p) \rangle &=& i f_{\eta_b} p_{\rho}.
\label{feta}
\eeq
At this point, some comments are in order.
\begin{enumerate}
\item
In eq.~(\ref{formfactor}) there is a clear separation of large- and small-distance 
contributions.
While $H(\xi_1,\xi_2, \mu)$ describes the hard part of the amplitude,
the large-distance part is parameterized 
by the combination of the DAs, which effectively include resummation of the relativistic
corrections to the amplitude. A discussion of this point can be found in
\cite{Braguta:2009df, Braguta:2008tg}.
%
\item
In eq.~(\ref{formfactor}) the dependence of the hard
part of the amplitude, the constants and the DAs on the scale $\mu$ is explicitly shown.
If the process in question 
were a leading-twist process, one could perform an exact
resummmation of all leading-twist radiative corrections to the amplitude,
$\sim \alpha_s \log(M_{\eta_b}^2/M_{J/\psi}^2)$,
simply by taking $\mu \sim M_{\eta_b}$ \cite{Chernyak:1983ej}.
Indeed, 
for a leading-twist process,
one would use the axial gauge, in which double-logarithmic and logarithmic corrections
only appear in the self-energy diagrams and re-scattering of final particles.
The double-logarithmic corrections are cancelled since final particles are colorless
objects, while the logarithmic corrections lead to the renormalization of the DAs
themselves.
Although the decay $\eta_b \to V_1 V_2$ is a next-to-next-to-leading-twist process,
all the arguments given above still seem to be applicable.
Note also that in eq.~(\ref{formfactor}) there is no divergence 
in the end-point region, $|\xi| \sim 1$, indicating that all
logarithms are collected.
These arguments allow us to believe that
eq.~(\ref{formfactor}) includes the exact resummation
of leading logarithmic radiative corrections to all loops.

\item
Whenever NRQCD and LC approaches are used to describe the same process, 
one should expect some kind of duality between the two results. 
For the process $\eta_b \to VV$ this duality can be checked 
at the leading-order approximation in relative velocity 
of the $c$-quark-antiquark pair inside charmonia. 
In particular, by taking infinitely narrow DAs and  
the constants $f_T, f_V$ and masses $M_V, 2 m_c$ at the next-to-leading order approximation 
in relative velocity \cite{Braguta:2007ge},
\beq
\frac {f_{T}} {f_V} = 1 - \frac {\langle v^2 \rangle } 3, \nonumber \\
\frac {M_V} {2 m_c} = 1 + \frac {\langle v^2 \rangle } 2,
\label{nrqcd_c}
\eeq
and by neglecting all radiative corrections, one gets from eq.~(\ref{formfactor}): 
\beq
F=\frac {256 \pi^2 \alpha_s^2} {81} \frac 1 {m_b^{6}} f_{\eta_b} f_V^2 m_c^2 \langle v^2 \rangle, 
\label{nrqcd}
\eeq
which coincides with the result obtained in \cite{Jia:2006rx}.
In these formulae, $\langle v^2 \rangle$ is the NRQCD matrix element, defined as 
\beq
\langle v^2 \rangle =- \frac 1 {m_c^2} 
\frac {\langle 0 | \chi^+ (\vec {\sigma} \vec {\epsilon }) ({\overset {\leftrightarrow} {\bf D} })^2 \varphi | V(\epsilon)  \rangle}
 {\langle 0 | \chi^+ (\vec {\sigma} \vec {\epsilon }) \varphi | V(\epsilon)  \rangle}.
\eeq
As noted in \cite{Braguta:2009df, Braguta:2008tg}, the duality 
between NRQCD and LC allows us to estimate the size of power corrections. 
The idea is that if one expands the NRQCD result in powers of $1/M_{\eta_b}$, than the first 
term coincides with the LC prediction and the second term
gives an estimate of power corrections to the LC result. Thus, power corrections 
to the amplitude of the  $\eta_b \to VV$ decay can be estimated 
as $\sim 4v^2 M_{V}^2/M_{\eta_b}^2$. 
\end{enumerate}
Now we have all the ingredients needed to calculate the rates of the decays $\eta_b \to V_1 V_2$.

\section{Numerical results and discussion}

\subsection{Input parameters}

In order to obtain numerical results for the branching ratios of the decays
$\eta_b \to J/\psi J/\psi, J/\psi \psi', \psi' \psi'$  the following input parameters were used:
\begin{enumerate}
\item
The strong coupling constant $\alpha_s ( \mu) $ is taken at the one loop, 
\beq
\alpha_s (\mu) = \frac {4 \pi} {\beta_0 \log (\mu^2/\Lambda^2)},
\eeq
with $\Lambda=0.2$ GeV, $\beta_0=25/3$. 
\item
The mass of  the $c$-quark in $\overline {MS}$ scheme, ${\overline m}_c=1.2$~GeV. 
\item
The leptonic decay constants of the $J/\psi$ and $\psi'$ mesons $f_{V}^{J/\psi}, f_{V}^{\psi'}$ 
were determined directly from experimental data, while the constants $f_{T}^{J/\psi}$ and $f_{T}^{\psi'}$ were calculated 
within NRQCD in \cite{Braguta:2007ge}: 
\beq
(f_{V}^{J/\psi} )^2 &=& 0.173 \pm 0.004~ \mbox{GeV}^2, \qquad~~~~~~ (f_{V}^{\psi'})^2 = 0.092 \pm 0.002~ \mbox{GeV}^2, \nonumber \\ 
(f_{T}^{J/\psi} (M_{J/\psi}) )^2 &=& 0.144 \pm 0.016~ \mbox{GeV}^2, \quad ( f_{T}^{\psi'} (M_{J/\psi}))^2 = 0.068 \pm 0.022~ \mbox{GeV}^2. 
\label{const1}
\eeq
\item
We assume that the total decay width of the $\eta_b$ meson $\Gamma_{\mbox{tot}}(\eta_b)$  can be approximated by its
two-gluon decay width $\Gamma(\eta_b \to gg)$ which, at the leading order  
in relative velocity and $\alpha_s$, is equal to
\beq
\Gamma_{\mbox{tot}}(\eta_b)=\Gamma(\eta_b \to gg) = \frac {8 \pi} {9} \frac {\alpha_s^2} {M_{\eta_b}} f_{\eta_b}^2\;. 
\eeq
\item
The leading twist DAs needed for the calculations are taken from models developed in
\cite{Braguta:2006wr, Braguta:2007fh, Braguta:2007tq, {Braguta:2008qe}}. 
\end{enumerate}

\subsection{Estimation of uncertainties}

The most important uncertainties come from the following sources:

\begin{enumerate}
\item
{\it Model-dependence of the DAs}. These uncertainties can be estimated by 
varying the parameters of these models 
(see \cite{Braguta:2006wr, Braguta:2007fh, Braguta:2007tq, {Braguta:2008qe}}  for more details).
The calculations show that for the processes $\eta_b \to J/\psi J/\psi, J/\psi \psi', \psi' \psi'$
these uncertainties are no larger than $\sim 5 \%,~ 13 \%,~ 30 \%$, respectively.
In fact, these uncertainties are expected to be rather low, due to the property
that the precision of any DA model improves with evolution \cite{Braguta:2006wr}. 

\item
{\it Radiative corrections}. 
Within the approach used in this paper, the leading logarithmic radiative corrections due to the 
evolution of the DAs and the strong coupling constant were effectively resummed.
Although we argued above that this is also true for   
all leading logarithmic radiative corrections, there is no strict 
proof of this statement. For this reason, we estimate 
the uncertainty due to the radiative corrections as  
$\sim \alpha_s( M_{\eta_b} /2 ) \log (M_{\eta_b}^2/(4 M_{J/\psi}^2) ) \sim 50 \%$. 

\item
{\it Power corrections.} 
As mentioned above,
this source of uncertainty can be estimated as $\sim 4 \langle v^2 \rangle M_{V}^2/M_{\eta_b}^2$, 
which is the largest for the decay 
$\eta_b \to \psi' \psi'$, reaching 
$\sim 4 \langle v^2 \rangle_{\psi'} M_{\psi'}^2/M_{\eta_b}^2 \sim 20 \%$.

\item
{\it Relativistic corrections.}  
This source of uncertainty appears because we treated $\eta_b$ meson at the 
leading-order approximation in relative velocity. It can be estimated 
as $\sim v_{\eta_b}^2 \sim 10 \%$.

\item
{\it The uncertainties in the values of constants (\ref{const1}).} 
For the three processes 
$\eta_b \to J/\psi J/\psi, J/\psi \psi', \psi' \psi'$ these errors are estimated to be
$\sim 16 \%, 27 \%,  49 \%$, respectively.

\item
{\it Higher Fock states}. It can be argued that
at the scale $\mu$ relevant to $\eta_b$ decay process, only a small fraction of 
quarkonium momentum is carried by the quark-gluon sea, typically $\sim 5-10\%$
\cite{Kartvelishvili:1985ac}.
Hence, we expect the effects of higher Fock states to be negligible, compared to
other uncertainties considered here. 

\end{enumerate}

The overall uncertainties of our calculations were obtained by adding the above errors in quadrature.

\subsection{Results and discussion}

By substituting the expressions for DAs and the necessary constants into 
eqs.~ (\ref{formfactor}) and (\ref{gamma}), we get the following values for
the three branching ratios:
\beq
Br(\eta_b \to J/\psi J/\psi) &=& (6.2 \pm 3.5 ) \times 10^{-7}, \nonumber \\ 
\label{results}
Br(\eta_b \to J/\psi \psi') &=& (10 \pm 6 ) \times 10^{-7},  \\ 
Br(\eta_b \to \psi' \psi') &=& (3.7 \pm 2.8 ) \times 10^{-7}.\nonumber
\eeq

It is interesting to compare these results with previous calculations.
In particular, within the leading order NRQCD, one has \cite{Jia:2006rx}:
\beq
Br(\eta_b \to J/\psi J/\psi) = (2.4^{+4.2}_{-1.9} ) \times 10^{-8}.
\eeq
which is roughly 20 times smaller than our result shown above.
The reason of this suppression can be traced to the 
expression for the amplitude (\ref{formfactor}),
where all terms are in fact proportional to the constants $f_A$ and $f_S$, 
which, in turn, are expressed through $f_V$ and $f_T$ (see eq.~(\ref{fafs})).
In the absence of relativistic and radiative corrections, the fine-tuning between $f_V$, $f_T$ and the
masses, clearly visible in eqs.~(\ref{nrqcd_c}), guerantees that $f_A$, $f_S$ and hence
the formfactor $F$ are proportional to $\langle v^2 \rangle$, which is small for
nonrelativistic systems. 
%
Taking relativistic and leading logarithmic radiative corrections to the constants $f_A$ and $ f_S$
into account breaks the fine tuning, thus leading to a considerable enhancement of the
branching ratio.
%
To illustrate the above argument numerically, 
we take an infinitely narrow approximation for the DAs, parameters with 
fine-tuning given by eqs.~(\ref{nrqcd_c}), and $\langle v^2 \rangle = 0.25$,
to obtain $Br(\eta_b \to J/\psi J/\psi)\simeq 2 \times 10^{-8}$,
in agreement the leading order NRQCD result \cite{Jia:2006rx}. 
Next, we take into account relativisitic and leading logarithmic 
radiative corrections to the constants $f_A$ and $f_S$, but 
still use an infinitely narrow approximation for the DAs. 
In this case fine-tuning is broken, and we get $\sim 3 \times 10^{-7}$,
and order-of-magnitude increase compared to the NRQCD value.
By including renormalization group evolution and relativistic motion into the DAs,
we get a further increase of the branching ratio by a  
factor $\sim 2$.
%

In \cite{Gong:2008ue} the authors took into account one-loop radiative corrections
and obtained
\beq
Br(\eta_b \to J/\psi J/\psi) = (2.1-18.6 ) \times 10^{-8}.
\eeq
Although this number seems to be compatible with ours shown 
in eq.~(\ref{results}), we do not believe that the two results are in agreement
with each other. In particular, the analytical form of the formfactor $F$ 
obtained in \cite{Gong:2008ue} contains logarithmic terms:
\beq
{\mbox{Re}} F \sim \frac {19} {32} \log^2 { \frac {M_{\eta_b}^2} {M_{J/\psi}^2} }+... \nonumber \\ 
{\mbox{Im}} F \sim \pi \frac {19} {16} \log { \frac {M_{\eta_b}^2} {M_{J/\psi}^2} }+...
\label{reim}
\eeq
In the LC approach used in our calculation, all double logarithms cancel
as the final partciles are colourless objects \cite{Lepage:1980fj}.
Moreover, there are only two reasons why 
a general QCD amplitude may contain large logarithms:
renormalization and collinear divergences \cite{Lepage:1980fj, Smilga:1978bq}. 
Clearly, the imaginary part of $F$ is 
not renormalized at one loop, hence
the large logarithm in eq.~(\ref{reim}) must be due to a collinear divergence.
However, it is known that collinear divergences can be factored out, and do not have 
an imaginary part \cite{Smilga:1978bq}. In light of these arguments, the result
obtained in \cite{Gong:2008ue} looks strange.

The authors of \cite{Gong:2008ue} believe that there is no need for
renormalization in their calculation of the radiative corrections,
since the counterterms are proportional
to the leading order contribution, which vanishes
at the leading order in both $\alpha_s$ and $v_c$. 
We do not think that this statement is correct, 
since the expansion is done in operators which are
not multiplicatively renormalizable. Therefore,
the ultraviolet divergences may arise at the leading 
order in $v_c$ due to the $v_c$-suppressed operators. 
This effect violates NRQCD velocity scaling rules, and 
is discussed in detail in \cite{Braguta:2008tg, Braguta:2006wr}.

%
Yet another estimate for the same branching ratio was obtained in
 \cite{Santorelli:2007xg},
 where the final-state interaction effects due to a different decay
 mechanism were taken into account, yielding
 \beq
 Br(\eta_b \to J/\psi J/\psi) = (0.5 \times 10^{-8}-1.2 \times 10^{-5} ).
 \eeq

In conclusion, we have calculated the branching fractions of the decays
$\eta_b \to J/\psi J/\psi, J/\psi \psi', \psi' \psi'$ in the framework of
the light cone formalism. The uncertainties of our calculation have also 
been assessed. Our results, presented in eqs.~(\ref{results}), are more 
than an order of magnitude larger than those obtained within NRQCD.

\begin{acknowledgments}
The authors thank A.K. Likhoded and A.V. Luchinsky for useful discussion.
This work was partially supported by Russian Foundation of Basic Research under grant 07-02-00417.
\end{acknowledgments}


\begin{thebibliography}{**}


\bibitem{:2008vj}
  B.~Aubert {\it et al.}  [BABAR Collaboration],
  Phys.\ Rev.\ Lett.\  {\bf 101}, 071801 (2008)
  [Erratum-ibid.\  {\bf 102}, 029901 (2009)]
  [arXiv:0807.1086 [hep-ex]].

\bibitem{Braaten:2000cm}
  E.~Braaten, S.~Fleming and A.~K.~Leibovich,
  Phys.\ Rev.\  D {\bf 63}, 094006 (2001)
  [arXiv:hep-ph/0008091].

\bibitem{Kartvelishvili:1984en}
  V.~G.~Kartvelishvili and A.~K.~Likhoded,
  Yad.\ Fiz.\  {\bf 40}, 1273 (1984).

\bibitem{Bodwin:1994jh}
  G.~T.~Bodwin, E.~Braaten and G.~P.~Lepage,
  Phys.\ Rev.\ D {\bf 51}, 1125 (1995)
  [Erratum-ibid.\ D {\bf 55}, 5853 (1997)]
  [arXiv:hep-ph/9407339].
 

\bibitem{Jia:2006rx}
  Y.~Jia,
  Phys.\ Rev.\  D {\bf 78}, 054003 (2008)
  [arXiv:hep-ph/0611130].


\bibitem{Gong:2008ue}
  B.~Gong, Y.~Jia and J.~X.~Wang,
  Phys.\ Lett.\  B {\bf 670}, 350 (2009)
  [arXiv:0808.1034 [hep-ph]].

\bibitem{Santorelli:2007xg}
  P.~Santorelli,
  Phys.\ Rev.\  D {\bf 77}, 074012 (2008)
  [arXiv:hep-ph/0703232].



\bibitem{Braaten:2002fi}
  E.~Braaten and J.~Lee,
  Phys.\ Rev.\ D {\bf 67}, 054007 (2003)
  [arXiv:hep-ph/0211085];

\bibitem{Liu:1}
K.~Y.~Liu, Z.~G.~He and K.~T.~Chao,
  Phys.\ Lett.\ B {\bf 557}, 45 (2003)
  [arXiv:hep-ph/0211181];

\bibitem{Liu:2}
  K.~Y.~Liu, Z.~G.~He and K.~T.~Chao,
  Phys.\ Rev.\  D {\bf 77}, 014002 (2008)
  [arXiv:hep-ph/0408141].

\bibitem{Zhang:2005ch}
  Y.~J.~Zhang, Y.~j.~Gao and K.~T.~Chao,
  Phys.\ Rev.\ Lett.\  {\bf 96}, 092001 (2006)
  [arXiv:hep-ph/0506076].

\bibitem{Gong:2007db}
  B.~Gong and J.~X.~Wang,
  Phys.\ Rev.\  D {\bf 77}, 054028 (2008)
  [arXiv:0712.4220 [hep-ph]].
 
\bibitem{Zhang:2008gp}
  Y.~J.~Zhang, Y.~Q.~Ma and K.~T.~Chao,
  Phys.\ Rev.\  D {\bf 78}, 054006 (2008)
  [arXiv:0802.3655 [hep-ph]].
 




\bibitem{Bondar:2004sv}
  A.~E.~Bondar and V.~L.~Chernyak,
  Phys.\ Lett.\ B {\bf 612}, 215 (2005)
  [arXiv:hep-ph/0412335].


\bibitem{Braguta:2005kr}
  V.~V.~Braguta, A.~K.~Likhoded and A.~V.~Luchinsky,
  Phys.\ Rev.\ D {\bf 72}, 074019 (2005)
  [arXiv:hep-ph/0507275].
  



\bibitem{Berezhnoy:2007sp}
  A.~V.~Berezhnoy,
  arXiv:hep-ph/0703143.

\bibitem{Ebert:2008kj}
  D.~Ebert, R.~N.~Faustov, V.~O.~Galkin and A.~P.~Martynenko,
  arXiv:0803.2124 [hep-ph].

\bibitem{He:2007te}
  Z.~G.~He, Y.~Fan and K.~T.~Chao,
  Phys.\ Rev.\  D {\bf 75}, 074011 (2007)
  [arXiv:hep-ph/0702239].



\bibitem{Bodwin:2007ga}
  G.~T.~Bodwin, J.~Lee and C.~Yu,
  Phys.\ Rev.\  D {\bf 77}, 094018 (2008)
  [arXiv:0710.0995 [hep-ph]].

\bibitem{Braguta:2008tg}
  V.~V.~Braguta,
  arXiv:0811.2640 [hep-ph].


\bibitem{Chernyak:1983ej}
  V.~L.~Chernyak and A.~R.~Zhitnitsky,
  Phys.\ Rept.\  {\bf 112}, 173 (1984).

\bibitem{Braguta:2009df}
  V.~V.~Braguta, A.~K.~Likhoded and A.~V.~Luchinsky,
  arXiv:0902.0459 [hep-ph].



\bibitem{Ball:1998sk}
  P.~Ball, V.~M.~Braun, Y.~Koike and K.~Tanaka,
  Nucl.\ Phys.\  B {\bf 529}, 323 (1998)
  [arXiv:hep-ph/9802299].






\bibitem{Braguta:2006wr}
  V.~V.~Braguta, A.~K.~Likhoded and A.~V.~Luchinsky,
  Phys.\ Lett.\  B {\bf 646}, 80 (2007)
  [arXiv:hep-ph/0611021].

\bibitem{Braguta:2007fh}
  V.~V.~Braguta,
  Phys.\ Rev.\  D {\bf 75}, 094016 (2007)
  [arXiv:hep-ph/0701234].


\bibitem{Braguta:2007tq}
  V.~V.~Braguta,
  Phys.\ Rev.\  D {\bf 77}, 034026 (2008)
  [arXiv:0709.3885[hep-ph]].

\bibitem{Braguta:2008qe}
  V.~V.~Braguta, A.~K.~Likhoded and A.~V.~Luchinsky,
  arXiv:0810.3607 [hep-ph].

\bibitem{Braguta:2007ge}
  V.~V.~Braguta,
  Phys.\ Rev.\  D {\bf 78}, 054025 (2008)
  [arXiv:0712.1475 [hep-ph]].



\bibitem{Kartvelishvili:1985ac}
  V.~G.~Kartvelishvili and A.~K.~Likhoded,
  Sov.\ J.\ Nucl.\ Phys.\  {\bf 42} (1985) 823.







\bibitem{Lepage:1980fj}
  G.~P.~Lepage and S.~J.~Brodsky,
  Phys.\ Rev.\ D {\bf 22}, 2157 (1980).


\bibitem{Smilga:1978bq}
  A.~V.~Smilga and M.~I.~Vysotsky,
  Nucl.\ Phys.\ B {\bf 150}, 173 (1979).






\end{thebibliography}
\end{document}